\documentclass[aps,prl,reprint,twocolumn,superscriptaddress,showpacs]{revtex4-2}

\usepackage{amsfonts}
\usepackage{amsmath}
\usepackage{graphicx}
\usepackage{bm}
\usepackage{amssymb}
\usepackage{dcolumn}
\usepackage{color}
\usepackage{multirow}
\usepackage{booktabs}
\usepackage[colorlinks,
linkcolor=blue,
anchorcolor=blue,
citecolor=blue,
urlcolor=blue,
]{hyperref}
\begin{document}
	\title{Large kagome family candidates with topological superconductivity and charge density waves}
	
	\author{Xin-Wei Yi}
	\affiliation{School of Physical Sciences, University of Chinese Academy of Sciences, Beijing 100049, China}
	\author{Xing-Yu Ma}
	\affiliation{School of Physical Sciences, University of Chinese Academy of Sciences, Beijing 100049, China}
	\author{Zhen Zhang}
	\affiliation{School of Physical Sciences, University of Chinese Academy of Sciences, Beijing 100049, China}
	
	\author{Zheng-Wei Liao}
	\affiliation{School of Physical Sciences, University of Chinese Academy of Sciences, Beijing 100049, China}
	
	\author{Jing-Yang You}
	\email{phyjyy@nus.edu.sg}
	\affiliation{Department of Physics, Faculty of Science, National University of Singapore, 117551, Singapore}
	
	\author{Gang Su}
	\email{gsu@ucas.ac.cn}
	\affiliation{Kavli Institute for Theoretical Sciences, and CAS Center for Excellence in Topological Quantum Computation, University of Chinese Academy of Sciences, Beijing 100190, China}
	\affiliation{School of Physical Sciences, University of Chinese Academy of Sciences, Beijing 100049, China}
	
	\begin{abstract}
		A group of newly discovered non-magnetic metal kagome structures AV$_3$Sb$_5$ (A=K, Rb, Cs) have aroused widespread interest in experiment and theory due to their unusual charge density wave (CDW) and intertwined superconductivity. However, they all possess weak electron-phonon coupling (EPC) and low superconducting transition temperature T$_c$. Here, we performed high-throughput first-principles calculations on novel kagome candidates with AV$_3$Sb$_5$ prototype structure, and proposed 24 dynamically novel stable kagome metals. The calculation based on Bardeen-Cooper-Schrieffer theory shows that most of these metals are superconductors with much stronger EPC than the reported AV$_3$Sb$_5$ materials, and their T$_c$ are between 0.3 and 5.0K. Additionally, several compounds, such as KZr$_3$Pb$_5$ with the highest T$_c$, are identified as $\mathbb{Z}$$_2$ topological metals with clear Dirac cone topological surface states near Fermi level, And NaZr$_3$As$_5$ is shown to have possible CDW phases. Our results provide rich platforms for exploring various new physics with kagome structure, in which the coexistence of superconductivity and nontrivial topological nature provides promising insights for the discovery of topological superconductors. 
	\end{abstract}
	\maketitle
	\emph{Introduction.---}In 2019, a new class of nonmagnetic metals AV$_3$Sb$_5$ (A =K, Rb, and Cs) with perfect vanadium kagome net were synthesized \cite{Ortiz2019}. Since then, surprises have emerged in the study of these structures. The electronic structures of AV$_3$Sb$_5$ show Dirac nodal lines, nontrivial $\mathbb{Z}$$_2$ topological band indices and clear topological surface states near Fermi level, indicating that their normal states are $\mathbb{Z}$$_2$ topological metals \cite{Ortiz2019,Ortiz2021,Ortiz2020,Zhao2021}. The superconducting transition temperatures T$_c$ of KV$_3$Sb$_5$, RbV$_3$Sb$_5$ and CsV$_3$Sb$_5$ are 0.93 \cite{Ortiz2021}, 0.92 \cite{Yin2021} and 2.5K \cite{Ortiz2020}(2.3K \cite{Liang2021,Chen2021}), respectively, and the temperatures T* consponding charge density wave (CDW) transition are 78 \cite{RN132}, 102 \cite{Li2021} and 94K \cite{Chen2021a}, respectively. The exotic CDW states of non electron-phonon coupling (EPC) mechanism \cite{Li2021} display many unconventional characteristics. The CDWs in these three compounds exhibit chiral anisotropy \cite{Jiang2021,Wang2021,Shumiya2021} and reduce the density of electronic states near the Fermi level \cite{Jiang2021,Uykur2022,Ratcliff2021,Cho2021}. Various evidences, including giant anomalous Hall response \cite{Yu2021,Yang2020}, CDW chirality adjustable by magnetic field \cite{Jiang2021,Wang2021,Shumiya2021}, edge supercurrent \cite{Wang2020} and spontaneous internal magnetic field \cite{Mielke2022}, indicate that the charge order may break the time-reversal symmetry, which has also been verified theoretically \cite{Jiang2021,Denner2021,Lin2021}. Moreover, CsV$_3$Sb$_5$ samples have been found to own roton pair density wave \cite{Chen2021}, which is similar to the one in unconventional high-T$_c$ cuprate superconductors \cite{Ruan2018}. The coexistence of V- and U-shaped superconducting gaps \cite{Xu2021} and the anisotropic superconducting properties in CsV$_3$Sb$_5$ \cite{Ni2021} imply the possible multi-band superconducting pairing. Intertwinded superconductivity with CDW shows many new features. For instance, T$_c$ exhibits an unconventional double dome behavior, and T* decreases rapidly with increasing pressure \cite{Chen2021a,Yu2021a,Du2021}. The second dome of T$_c$ and the disappearance of T* occur at the same pressure. The superconducting properties and charge order of AV$_3$Sb$_5$ can also be tuned by magnetic impurity \cite{Xu2021}, strain \cite{Yin2021a} and thickness \cite{Wang2021a,Song2021,Song2021a}, which dramatically enriches the phase diagram. The experimental and theoretical studies on AV$_3$Sb$_5$ show a complementary and rapid trend. However, to further explore the exotic properties of AV$_3$Sb$_5$, more candidate compounds based on the AV$_3$Sb$_5$ prototype structure are urgently needed.
	
	In this paper, we first apply the high-throughput first-principles calculations to 800 new kagome structures based on AV$_3$Sb$_5$ prototype, and find 24 dynamically stable metal compounds. Then, we carefully study their superconducting and topological properties. The results show that 14 novel compounds are superconductors with the T$_c$ between 0.3 and 5.0K. Moreover, several structures, including KZr$_3$Pb$_5$ with the highest T$_c$, have strong $\mathbb{Z}$$_2$ indices with abundant nontrivial topological surface states near Fermi surface, suggesting that they are $\mathbb{Z}$$_2$ topological metals. The coexistence of superconductivity and nontrival band topology opens a door for the discovery of topological superconductivity based on the kagome net. Additionally, we also find two possible CDW phases in NaZr$_3$As$_5$, which exhibit soft modes in phonon spectrum, and may provide useful information for further understanding the CDW phases in AV$_3$Sb$_5$.
	
\begin{figure}[t]
		\centering
		\includegraphics[scale=0.7,angle=0]{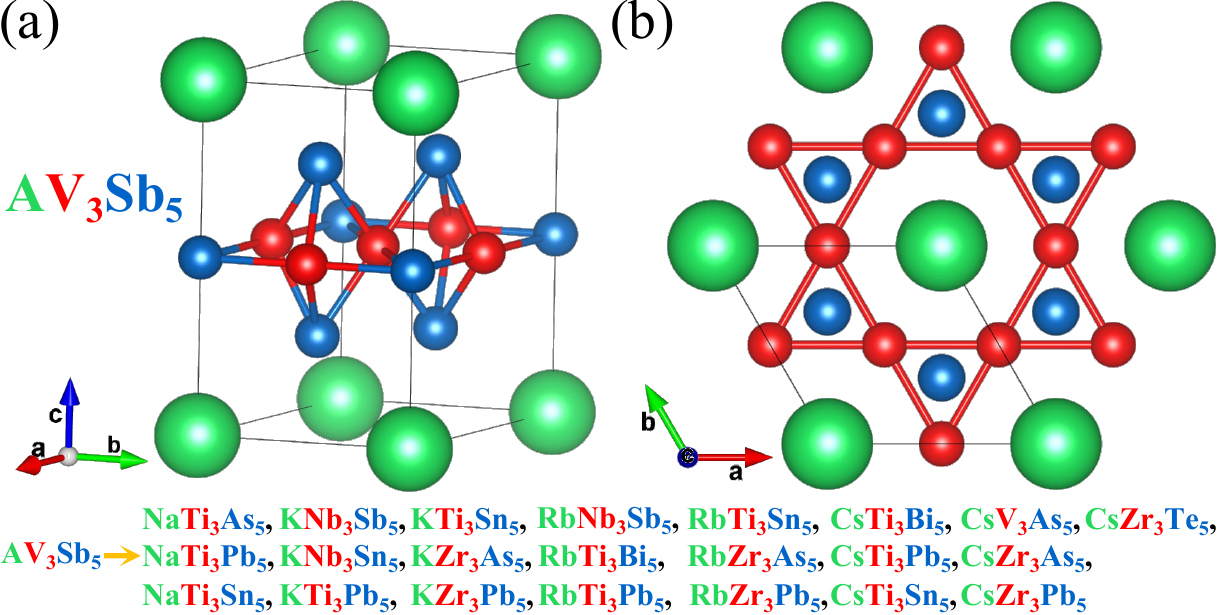}\\
		\caption{The crystal structure of AV$_3$Sb$_5$. 22 new stable AB$_3$C$_5$ members with the same crystal structure as AV$_3$Sb$_5$ are also indicated.}\label{fig1}
	\end{figure}

	\emph{Crystal structure of $AV_3Sb_5$.---}The AV$_3$Sb$_5$ crystallize in a layered structure with the space group of P6/mmm (No.191) as shown in FIG.~\ref{fig1}. The perfect V-kagome net mixed with the Sb-triangular net of is located in the middle layer, which is sandwiched by two additional honeycomb layers of Sb atoms. The upper and lower triangular layers of alkali metal A atoms have a large bond distance relative to the middle V-Sb layer and are loosely bonded to them. 
	
	\emph{Searching new structures.---}The high-throughput first-principles calculations are used to search for kagome topological superconductor candidates (as indicated in FIG. S1 in Supplementary Material (SM)). Based on the prototype structure of AV$_3$Sb$_5$, 800 new compounds are constructed by replacing A with alkali metal elements Li, Na, K, Rb and Cs, replacing V with all transition metal elements in the fourth and fifth periods of the periodic table, and replacing Sb with its neighboring elements (Ge, As, Se, Sn, Sb, Te, Pb, and Bi). These new compounds will be abbreviated as AB$_3$C$_5$ below. For all these new compounds, we first carry out fully geometric relaxation in different magnetic configurations (see FIG. S2). Then, by comparing the total energies of different magnetic configurations, each AB$_3$C$_5$ member can be classified as nonmagnet (NM), ferromagnet (FM) and antiferromagnet (AFM). Phonon spectra are calculated to determine the dynamic stability of these AB$_3$C$_5$ members. Compounds without imaginary frequency in phonon spectra will be further analyzed for their corresponding electronic structures, superconductivity, and topological properties. On the other hand, compounds with imaginary phonon frequency will be used to discuss possible CDW phase.
	
	In doing so, we discover 24 novel stable AB$_3$C$_5$ members, including 22 NM, one FM (CsTi$_3$Pb$_5$) and one AFM (RbCr$_3$Te$_5$), and their lattice information is listed in Table S1. The magnetic calculations show that CsTi$_3 $Pb$_5 $ has a ferromagnetic ground state with a magnetic easy-axis anisotropy and the magnetocrystalline anisotropy energy is -1.59meV per formula unit. RbCr$_3 $Te$_5$ has an in-plane 120° AFM ground state. The Monte Carlo simulations estimate that the Curie temperature of CsTi$_3 $Pb$_5 $ is about 18K, and the Neel temperature of RbCr$_3 $Te$_5$ is about 13K. In addition, we also calculate the formation energies of these compounds to further demonstrate their stabilities, as listed in Table S1. Electronic structure calculations show that they are all metals similar to AV$_3$Sb$_5$. Furthermore, 22 members maintain the same crystal structure as AV$_3$Sb$_5$ after structural optimization (as listed in FIG.~\ref{fig1}), except for CsRu$_3$Ge$_5$ and RbCr$_3$Te$_5$, whose triangles in the kagome nets are twisted, which changes their space group to P$\overline{6}$2m. 
	\begin{table}[t]
		\renewcommand\arraystretch{1.15}
		\caption{Electronic density of states at Fermi energy N(E$_F$) for per fomular unit (eV$^{-1}$f.u.$^{-1}$), logarithmic average frequency $\omega$$_l$$_o$$_g$ (K), EPC $\lambda$($\omega$ = $\infty$) and T$_c$ of 14 stable compounds.}
		{\centering
			\begin{tabular}{lp{1.7cm}<{\centering}p{1.7cm}<{\centering}p{1.3cm}<{\centering}p{1.3cm}<{\centering}p{1.3cm}}
				\hline
				\hline
				& N(E$_F$) (eV$^{-1}$f.u.$^{-1}$)  & $\omega$$_l$$_o$$_g$  (K)  & $\lambda$ & T$_c$ (K)\\
				\hline
				KNb$_3$Sn$_5$      &4.22   &149.1   &0.52     &2.102\\
				CsRu$_3$Ge$_5$      &3.19   &170.8   &0.36    &0.353\\
				RbTi$_3$Bi$_5$      &5.96   &149.8   &0.41    &0.719\\     
				CsTi$_3$Bi$_5$      &5.96   &163.4   &0.35    &0.316\\	
				KTi$_3$Pb$_5$       &7.29    &157.9   &0.51    &2.039\\			
				RbTi$_3$Pb$_5$      &7.50   &156.5   &0.50    &1.857\\			
				KTi$_3$Sn$_5$      &6.57    &180.3   &0.42    &0.974\\			
				RbTi$_3$Sn$_5$      &6.50   &182.5   &0.42    &0.961\\			
				CsTi$_3$Sn$_5$      &6.87   &174.6   &0.45    &1.375\\			
				CsZr$_3$As$_5$       &3.40   &125.8   &0.56    &2.289\\		
				KZr$_3$Pb$_5$       &6.47   &94.1   &0.91    &5.027\\			RbZr$_3$Pb$_5$      &6.56   &111.9   &0.72    &4.154\\			CsZr$_3$Pb$_5$     &6.53   &119.2   &0.58    &2.438\\			CsZr$_3$Te$_5$      &2.88   &123.5   &0.48 &1.266\\							
				\hline
				\hline
		\end{tabular}}
		\label{Table1}
	\end{table}
	
	\emph{Superconductivity.---}The metallicity of these 24 AB$_3$C$_5$ members enables us to perform further calculations on superconductivity. We calculate the Eliashberg spectral function $\alpha$$^2$F($\omega$) and cumulative EPC $\lambda$($\omega$) of these materials at ambient pressure, and estimate the T$_c$ with the McMillan-Allen-Dynes approach of Bardeen-Cooper-Schrieffer (BCS) theory (See SM for details). After careful calculation and screening, we found that 14 of these 24 members have the superconducting ground state as listed in Table~\ref{Table1}, where KZr$_3$Pb$_5$ possesses the highest T$_c$ of 5.027K, which is more than twice the experimental values of AV$_3$Sb$_5$ (see Table~\ref{Table2}) \cite{Ortiz2021,Ortiz2020,Chen2021,Liang2021,Yin2021}.
	
	\begin{table}[hb]
		\renewcommand\arraystretch{1.15}
		\caption{Electronic density of states at Fermi energy N(E$_F$) (eV$^{-1}$f.u.$^{-1}$), EPC $\lambda$($\omega$ = $\infty$), calculated T$_c$ and experimental superconducting temperature T$_c^{exp}$ for the ISD phase of AV$_3$Sb$_5$ \cite{Tan2021,Ortiz2021,Ortiz2021,Ortiz2020,Liang2021,Yin2021,Chen2021} and pristine phase of AZr$_3$Pb$_5$.}
		{\centering
			\begin{tabular}{lp{1.6cm}<{\centering}p{1.6cm}<{\centering}p{1.3cm}<{\centering}p{1.3cm}<{\centering}p{1.3cm}}
				\hline
				\hline
				& N(E$_F$) (eV$^{-1}$f.u.$^{-1}$)  & $\lambda$ & T$_c$ (K) &T$^{exp}_c$ (K)\\
				\hline
				KV$_3$Sb$_5$   &2.9   & 0.38    &0.22  	&0.93\\
				RbV$_3$Sb$_5$  &2.33   &0.32    &0.05		&0.92\\
				CsV$_3$Sb$_5$  &1.30   &0.25    &0.0008		&2.5 (2.3)\\				
				KZr$_3$Pb$_5$   &6.47   &0.91    &5.027  	&-\\
				RbZr$_3$Pb$_5$  &6.56   &0.72    &4.154		&-\\
				CsZr$_3$Pb$_5$  &6.53   &0.58    &2.438 &-\\							
				\hline
				\hline
		\end{tabular}}
		\label{Table2}
	\end{table}
	\begin{figure}[t]
		\centering
		\includegraphics[scale=0.252,angle=0]{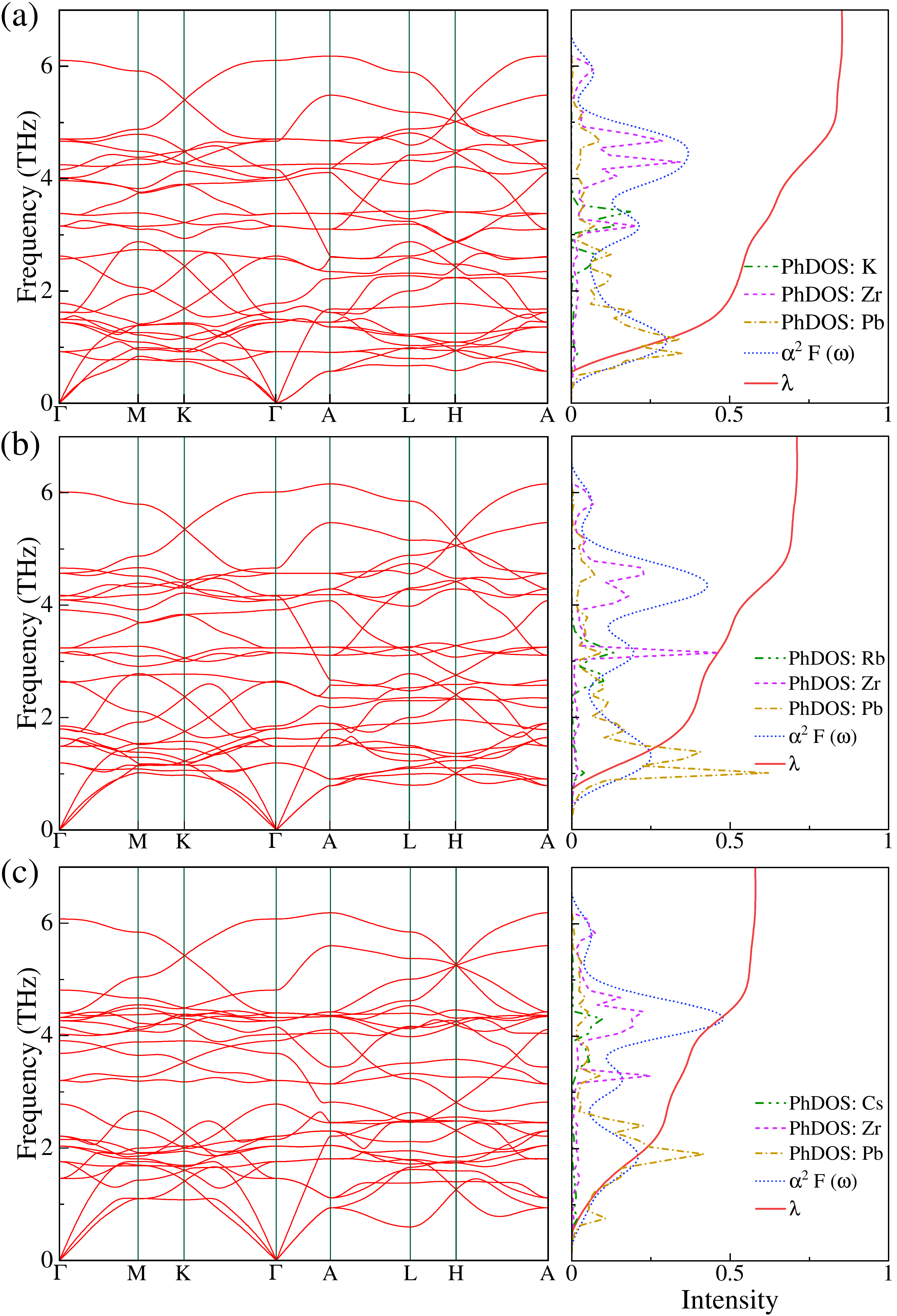}\\
		\caption{The phonon spectra, projected PhDOS, Eliashberg spectral function $\alpha$$^2$F($\omega$), and cumulative frequency dependent EPC $\lambda$($\omega$) of (a) KZr$_3$Pb$_5$, (b) RbZr$_3$Pb$_5$, (c) CsZr$_3$Pb$_5$.}\label{fig2}
	\end{figure}
	
	We choose AZr$_3$Pb$_5$ group with relative higher T$_c$ for further discussions. The phonon spectra, phonon density of states (PhDOS), $\alpha$$^2$F($\omega$) and $\lambda$($\omega$) of this group are plotted in FIG.~\ref{fig2}. We can see that the phonon spectra of three compounds are very similar. Careful comparison of their phonon spectra shows that the faint phonon softening at L point gradually becomes obvious from K to Rb to Cs. It can be seen from PhDOS that the contributions of Pb and Zr atoms to PhDOS are mainly distributed in the relatively low and high frequency regions with much prominent peaks, respectivley, while the PhDOS of alkali metal atoms distributed in the medium frequency region are very small. The relatively low frequency ($<$3THz) phonons corresponding to the vibration modes of Pb account for more than half of the total EPC. The T$_c$ of the three compounds decreases with the gradual increase of the atomic number of alkali metals as shown in Table~\ref{Table1}. The gradual decrease of T$_c$ from K to Rb to Cs is due to the negligible contribution of alkali metal elements to the EPC, and the increase of atomic radius from K to Cs, resulting in the gradual increase of the lattice parameters, is equivalent to applying a negative pressure (tensile strain) to the lattice, which significantly reduce the parameters related to the lattice and weakens the EPC.

\begin{figure}[b]
		\centering
		\includegraphics[scale=0.243,angle=0]{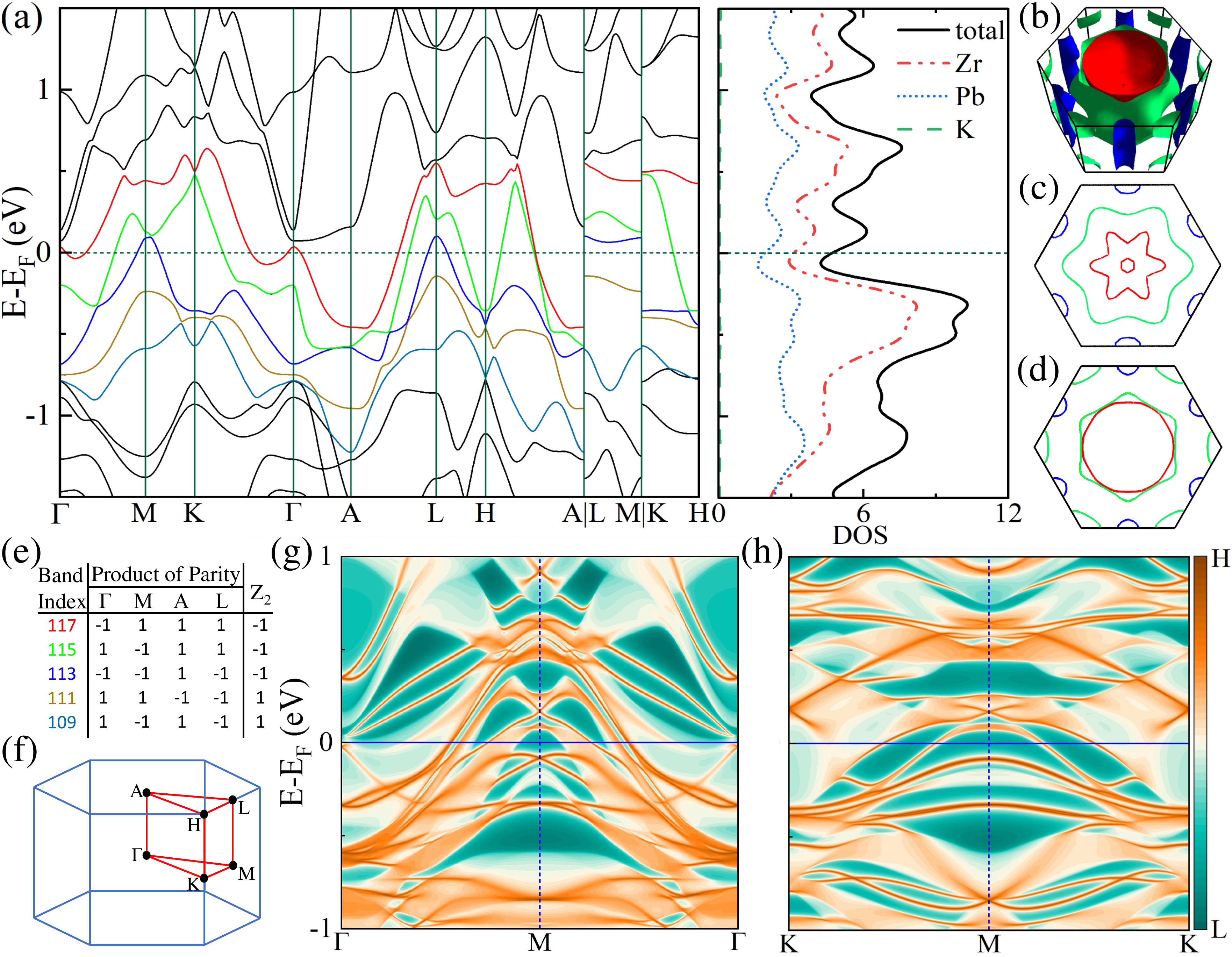}\\
		\caption{(a) The electronic energy bands and density of states calculated with SOC for KZr$_3$Pb$_5$. (b) 3D FS of KZr$_3$Pb$_5$, and its 2D maps at (c) k$_z$= 0 and (d) $\pi$ slices. Different colors of FS refers to different band indices consistent with (a). (e) Product of parity and $\mathbb{Z}$$_2$ indices of bands near Fermi level. (f) The Brillouin zone with high symmetry paths indicated. Topological surface states along (g) $\Gamma$-M-$\Gamma$ and (h) K-M-K paths on (001) plane for KZr$_3$Pb$_5$.}\label{fig3}
	\end{figure}

	\emph{Electronic band structure and topological property.---}We plot the electronic energy bands and density of states (DOS) with spin-orbit coupling (SOC) for KZr$_3$Pb$_5$ in FIG.~\ref{fig3}(a). The electronic band structures of RbZr$_3$Pb$_5$ and CsZr$_3$Pb$_5$ are also given in FIGs. S21-22. The 3D Fermi surface (FS) of KZr$_3$Pb$_5$ and its 2D slice at k$_z$=0 and $\pi$ planes are drawn in FIGs.~\ref{fig3}(b)-(d), which is obviously different from the FS of AV$_3$Sb$_5$ that exhibits strong 2D characteristics. Furthermore, we can see the obvious Fermi surface nesting with the nesting vector parallel to A-L and A-H in the k$_z$=$\pi$ slice.
	
	AZr$_3$Pb$_5$ are nonmagnetic similar to AV$_3$Sb$_5$. Therefore, combining the time-reversal and inversion symmetries of AZr$_3$Pb$_5$, we can get their $\mathbb{Z}$$_2$ topological invariants by calculating the parity of the wave functions at all time-reversal invariant momenta (TRIM) points \cite{Fu2007}. It can be seen from FIG.~\ref{fig3}(e) that all the energy bands passing through the Fermi surface have strong topological $\mathbb{Z}$$_2$ indices, resulting in abundant Dirac cone surface states near the Fermi level as shown in FIGs.~\ref{fig3} (g) and (h). The continuous bandgap between two energy bands in the whole Brillouin zone, nontrivial topological surface states and strong $\mathbb{Z}$$_2$ indices of bands near the Fermi level make KZr$_3$Pb$_5$ a $\mathbb{Z}$$_2$ topological metal. Similar analysis shows that many other AB$_3$C$_5$ members, such as CsZr$_3$Pb$_5$, are also $\mathbb{Z}$$_2$ topological metals.
	\begin{figure}[b]
		\centering
		\includegraphics[scale=0.171,angle=0]{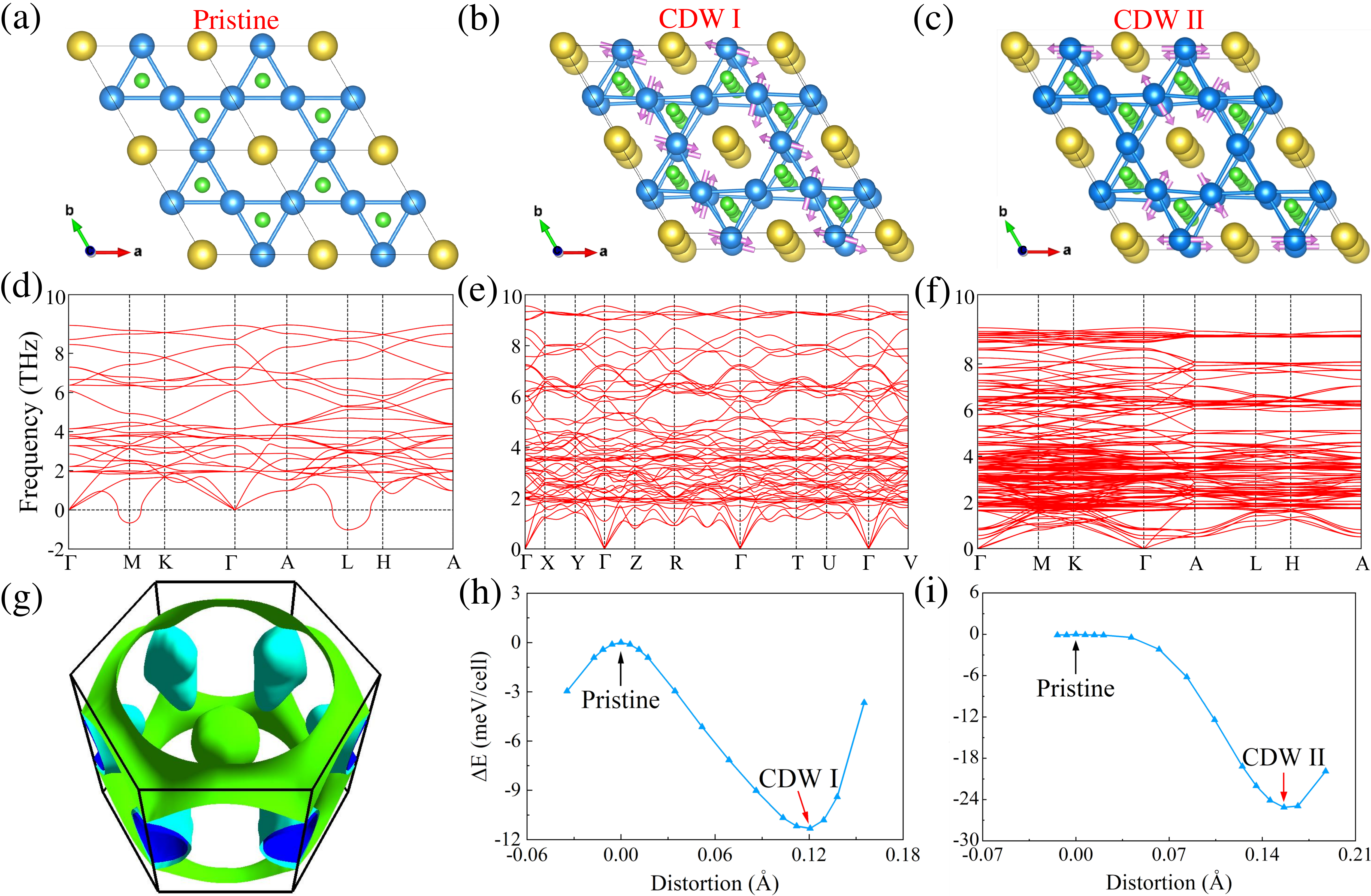}\\
		\caption{Crystal structures of NaZr$_3$As$_5$ in the (a) 2 × 2 supercell of pristine phase, (b) CDW I phase, and (c) the CDW II phase and their corresponding phonon spectra (d), (e) and (f), respectively. (g) 3D FS of NaZr$_3$As$_5$. The comparision of total energies $\Delta$E for (h) pristine phase and CDW I, (i) pristine phase and CDW II, where the distortion represents the displacement of Zr atoms and $\Delta$E stands for the relative total energy with respect to the pristine phase per cell with  72 atoms.The Brillouin zone of (b) and high symmetry paths of (e) are plotted in FIG. S30(d).}\label{fig4}
	\end{figure}
	
	\emph{Possible CDW phases.---}We take NaZr$_3$As$_5$ as an example to explore the possible CDW phases. One may find that the phonon spectrum in FIG.~\ref{fig4}(d) shows an obvious softening acoustic phonon modes at M and L points at the boundaries of Brillouin zone, and the imaginary frequency at L point is slightly larger than that at M point, which is very similar to AV$_3$Sb$_5$ \cite{Tan2021,Ratcliff2021}. The symmetry analysis on AV$_3$Sb$_5$ indicates that the irreducible representations of the imaginary mode at M and L points are M$_1^+$ and L$_2^-$, respectively, which is also consistent with the previous studies \cite{Ratcliff2021,Ortiz2021a}. However, similar analysis shows that the irreducible representations of the imaginary mode in NaZr$_3$As$_5$ are M$_2^+$ and L$_1^-$, which makes NaZr$_3$As$_5$ exhibit completely different distortions from AV$_3$Sb$_5$. In consideration of one L point, it gives the phase as shown in FIG.~\ref{fig4}(b). The soft mode at M point makes corner-shared triangles in layers rotate around the corner, while the soft mode at L point leads to the distortion of adjacent layers with an additional $\pi$-shift. Clockwise and counterclockwise distortions will generate the same structure. The combination of all three unequal L points gives a similar phase as shown in FIG.~\ref{fig4}(c), which differs from FIG.~\ref{fig4}(b) in that Zr atoms in the kagome layers rotate around the center of triangles. These two structures have Ibam (No.72) and P6/mcc (No.192) space groups, respectively. Both of them reduce the rotation symmetry of C$_6$ to C$_2$, but still retain the spatial inversion symmetry.
	
	Their phonon spectra in FIGs.~\ref{fig4}(e) and (f) show dynamic stability with completely disappeared imaginary frequency, so both of them are possible CDW phases of NaZr$_3$As$_5$. We label them CDW I and CDW II, respectively. Compared with the pristine phase, the displacement values of Zr atoms in kagome layer in CDW I and CDW II are 0.12 and 0.15\AA, respectively. The total energy as a function of displacement of Zr atoms is shown in FIGs.~\ref{fig4}(h) and (i). The total energies of the two stable CDW phases are 11.3 and 25.1meV lower than that of pristine structure, respectively. 
	
	Unfolded energy bands and DOS of CDW phases in FIG. S30 show no clear changes compared with the pristine phase except that some gaps are opened and the saddle point at the L point moves closer to the Fermi level. Saddle-point nesting in electronic structure is unlikely the origin of the CDW order NaZr$_3 $As$_5$. The real CDW phase, its origin and the possible interplay between charge order and superconductivity in NaZr$_3$As$_5$ deserve future experimental exploration. Besides NaZr$_3$As$_5$, we also plot those structures with obvious soft modes at high symmetry paths that may have CDW phases in FIG. S31.

	\emph{Discussion.---}In addition to AZr$_3$Pb$_5$, the calculated results of all other stable AB$_3$C$_5$ members are presented in FIGs. S3-23. The new AV$_3$C$_5$ members are not only structurally similar to AV$_3$Sb$_5$, but also inherit many attractive features, such as Van Hove singularities at high symmetry points near the Fermi level, Dirac points at the Fermi level, Dirac nodal lines, and strong 2D characteristics of the phonon spectrum and FS, which are worthy of further studies. 
	
	For all AB$_3$C$_5$ kagome families proposed in this paper and the reported AV$_3$Sb$_5$, we hardly see some obvious flat bands in band structures. To further interpret this feature, we construct a tight-binding model in SM. By tuning the hopping parameters, we find that with the increase of the hopping parameters between B atoms in the kagome lattice and C atoms, the flat band becomes more dispersive as seen in FIG. S27. C atoms and kagome B atoms are very close to each other, and the overlap of their orbitals makes the interaction between them very complex and destroys the destructive interference condition for the formation of a flat band in kagome lattice, resulting in the disappearance of the flat band.
	
	An important feature of those predicted structures beyond AV$_3$Sb$_5$ is their much stronger EPC strengths. The calculated T$_c$ of KV$_3$Sb$_5$, RbV$_3$Sb$_5$, and CsV$_3$Sb$_5$ based on BCS theory are 0.0008, 0.05 and 0.22K, respectively \cite{Tan2021}, which are much lower than their experimental values (see Table~\ref{Table2}), because the CDW in AV$_3$Sb$_5$ reduces the DOS near Fermi level and suppresses BCS superconductivity. This indicates there may be an unconventional superconducting mechanism. This mechanism is also expected to appear in the materials listed in Table~\ref{Table1}. From Table~\ref{Table2}, it can be observed that the calculated T$_c$ of AZr$_3$Pb$_5$ are much higher than those of AV$_3$Sb$_5$, thereby experimental T$_c$ of AZr$_3$Pb$_5$ may be higher.
	
	The coexistence of superconductivity and topological nontrivial surface states is essentially rare \cite{Fu2010,Wang2015,Sato2013,You2021}. It is reported that the robust zero-bias conductance peak in CsV$_3$Sb$_5$ exhibits similar characteristics to the Bi$_2$Te$_3$/NbSe$_2$ heterostructures with Majorana bound state \cite{Liang2021}. Our new compounds with both the superconducting ground state and the nontrivial topological surface states near the Fermi surface would provide a rich platform for exploring topological superconductivity and Majorana zero-energy modes.
	
	Mature experimental methods like flux method have been used to synthesize high-quality and stable AV$_3$Sb$_5$ compounds, which is a prerequisite for the rapid development of experimental analysis. In the initial work of Brenden \emph{et al}. for the AV$_3$Sb$_5$ family, they explore the combination of (K, Rb, Cs)(V, Nb, Ta)(Sb, Bi) under different synthetic conditions \cite{Ortiz2019}. However, only KV$_3$Sb$_5$, RbV$_3$Sb$_5$, and CsV$_3$Sb$_5$ are crystallized. In this work, 800 AB$_3$C$_5$ members in the high-throughput screening process  contain most of the combinations they explored. Our calculation results show that those compounds not synthesized in their experiment are dynamically unstable except KNb$_3$Sb$_5$. The agreement with the experimental results indicates that our present calculations are resonable, and the stable structures presented here are very likely to be synthesized in future experiments. Very recently, a newly discovered family of kagome metals RV$_6$Sn$_6$ (R = Gd, Ho, Y) with two V-derived kagome layers in the primitive cell was also synthesized by flux method \cite{Peng2021,Pokharel2021}. Therefore, the versatile and matured flux method may be employed to synthesize the stable structures in Table S1.
	
	\emph{Summary.---}In conclusion, we calculate 800 new kagome candidates based on the prototype structure of AV$_3$Sb$_5$ using a high-throughput DFT screening process, and discover 24 dynamically novel stable metal compounds, including one ferromagnetic, one antiferromagnetic and 22 nonmagnetic structures. These compounds display many appealing properties similar to AV$_3$Sb$_5$. Furthermore, based on the McMillan-Allen Dynes approach, 14 compounds among them are predicted to be phonon-mediated BCS superconductors with T$_c$ between 0.3-5K. KZr$_3$Pb$_5$ with the highest T$_c$ exhibits strong $\mathbb{Z}$$_2$ invariants of the energy bands and abundant nontrivial topological surface states near the Fermi level, revealing that it is a $\mathbb{Z}$$_2$ topological metal. In addition, we also find two possible CDW phases in NaZr$_3$As$_5$. This present work would give more insights on the exploration of possible topological superconductors.
	\section* {Acknowledgments}
	This work is supported in part by the National Key R\&D Program of China (Grant No. 2018YFA0305800), the Strategic Priority Research Program of the Chinese Academy of Sciences (Grants No. XDB28000000), the National Natural Science Foundation of China (Grant No.11834014), and High-magnetic field center of Chinese Academy of Sciences.

%

\end{document}